\documentclass[12pt,preprint]{aastex}

\newcommand\lsim{\mathrel{\rlap{\lower4pt\hbox{\hskip1pt$\sim$}}
        \raise1pt\hbox{$<$}}}
\newcommand\gsim{\mathrel{\rlap{\lower4pt\hbox{\hskip1pt$\sim$}}
        \raise1pt\hbox{$>$}}}

\newcommand\bb[1] {   \mbox{\boldmath{$#1$}}  }

\newcommand\del{\bb{\nabla}}
\newcommand\bcdot{\bb{\cdot}}

\newcommand\vv{\bb{v}}
\newcommand\B{\bb{B}}
\newcommand\BV{Brunt-V\"ais\"al\"a\ }

\newcommand\kva{ \bb{k\cdot v_A}  }

    \def\dd{\partial}
    \def\tilde{\widetilde}

    \def\beq{ \begin{equation} }
    \def\eeq{ \end{equation} }
    \def\spose#1{\hbox to 0pt{#1\hss}} % from Scott Tremaine
    \def\ltsim{\mathrel{\spose{\lower.5ex\hbox{$\mathchar"218$}}
         \raise.4ex\hbox{$\mathchar"13C$}}}

\def\tilde{\widetilde}

\long\def\Ignore#1{\relax}

\shorttitle{Transport in the Early Sun}
\shortauthors{Menou \& Le Mer}

\begin{document}

\title{Magneto-Rotational Transport in the Early Sun}

\author{Kristen Menou}
\affil{Department of Astronomy, Columbia University, 550 West
120th Street, New York, NY 10027, USA}
\and
\author{Jo\"{e}l Le Mer}
\affil{Ecole Polytechnique, Route de Saclay, 91128 Palaiseau, France
\& \\ Department of Astronomy, Columbia University, 550 West 120th
Street, New York, NY 10027, USA}

\begin{abstract}
Angular momentum transport must have occurred in the Sun's radiative
zone to explain its current solid body rotation.  We survey the
stability of the early Sun's radiative zone with respect to diffusive
rotational instabilities, for a variety of plausible past
configurations. We find that the (faster rotating) early Sun was prone
to rotational instabilities even if only weak levels of radial
differential rotation were present, while the current Sun is
not. Stability domains are determined by approximate balance between
dynamical and diffusive timescales, allowing generalizations to other
stellar contexts. Depending on the strength and geometry of the weak
magnetic field present, the fastest growing unstable mode can be
hydrodynamic or magneto-hydrodynamic (MHD) in nature.  Our results
suggest that diffusive MHD modes may be more efficient at transporting
angular momentum than their hydrodynamic
(``Goldreich-Schubert-Fricke'') counterparts because the minimum
spatial scale required for magnetic tension to be destabilizing limits
the otherwise very small scales favored by double-diffusive
instabilities.  Diffusive magneto-rotational instabilities are thus
attractive candidates for angular momentum transport in the early
Sun's radiative zone.

\end{abstract}

\keywords{hydrodynamics -- instabilities -- stars: rotation -- Sun:
rotation -- turbulence}

\section{Introduction} \label{sec:intro}

The Sun's internal structure, which has been probed extensively with
seismological techniques (Thompson et al. 2003), is a unique
laboratory for the study of fluid dynamical processes on astronomical
scales. In this context, it is rather significant that the physical
process at the origin of the solid body rotation in the Sun's
radiative zone has not yet been unambiguously identified. By analogy
with young Sun-like stars, it is believed that the young Sun was
rotating faster than it currently is (by a factor up to $\sim 50$) and
that it was subsequently spun down by magnetic torques associated with
the solar wind (see, e.g., Sofia et al. 1991 for a review).  Although
details are missing, the end result of this rotational evolution
should be a differentially rotating interior, with a faster rotating
core. Consequently, the well-established solid body rotation in the
current Sun's radiative zone is evidence that an efficient angular
momentum transport mechanism has operated in the past. It is also
tempting to associate this transport mechanism with ``deep,''
large-scale mixing of elements (below the convection zone), as
required to explain the much depleted lithium abundance at the solar
photosphere (e.g. Chaboyer, Demarque \& Pinsonneault 1995a, 1995b).

The motivation to identify a successful transport mechanism for the
Sun's radiative zone goes in fact much beyond the Sun itself, as this
mechanism is likely to also have consequences for the evolution of
rotating massive stars (e.g., Woosley, Heger \& Weaver 2002), the spin
of newly-formed compact objects resulting from this evolution
(e.g. Ott et al. 2006) and possibly the origin of gamma-ray bursts
(e.g. MacFadyen \& Woosley 1999). The mechanism could be operating in
the radiative zones of all kinds of stars or only at some stages of
their evolution, but until its nature is known, it is difficult to be
more specific.

Efforts to understand transport in stellar radiative zones are not new
(e.g. Tassoul 1978). It was clear early on that the microscopic
kinematic viscosity in the Sun, $\nu_\odot$, would not be sufficient
to reduce any global scale differential rotation present, since the
corresponding viscous timescale,
\begin{equation} \label{eq:tvis}
 \tau_{\rm vis} \simeq \frac{R_\odot^2}{\nu_\odot} \sim
 10^{12-13}~{\rm years},
\end{equation}
is prohibitively long (e.g. Goldreich \& Schubert 1967). To explain
the solid body rotation of the solar radiative zone which was later
revealed by helioseismology, another process must have been operating.

Several candidate mechanisms have indeed been proposed. The list
includes the so-called GSF instability (Goldreich \& Schubert 1967;
Fricke 1968), the ``secular hydrodynamical shear instability'' (Zahn
1974) and internal gravity waves (e.g. Kumar \& Quataert 1997; Talon
\& Zahn 1998; Talon, Kumar \& Zahn 2002; Charbonnel \& Talon 2005). No
consensus has been reached yet as to which of these mechanisms, if
any, provides a satisfactory answer to the current state of solid body
rotation. The GSF instability alone is not expected to bring a
radiative zone to a state of solid body rotation because its marginal
stability conditions still allow for substantial differential rotation
across cylindrical radii (Goldreich \& Schubert 1967). The existence
of the secular shear instability has not been rigorously established
but rather suggested on the basis of heuristic arguments (Zahn 1974;
Schatzman 1991), which makes it a difficult subject of detailed
study. The efficiency of internal gravity waves at transporting
angular momentum has been much debated in the literature (e.g. Press
1981; Talon \& Zahn 1998; Talon, Kumar \& Zahn 2002; Charbonnel \&
Talon 2005; Rogers \& Glatzmeier 2005, 2006). It is also a priori
unclear whether waves, which by nature transport energy and momentum
but not the fluid itself, can achieve a sufficient level of deep
mixing of elements such as lithium, whose depleted photospheric
abundance is arguably more easily attributed to instabilities and
turbulent mixing (but see Garcia-Lopez \& Spruit 1991; Schatzman 1993;
Charbonnel \& Talon 2005 for a different view).

In recent years, magnetic fields have taken central stage in this
discussion (e.g. Turck-Chieze et al. 2005). The role of a magnetic
field has been discussed in relation to that of internal gravity waves
(e.g. Mathis \& Zahn 2005; Rashba, Semikoz \& Valle 2005). Solar
evolutionary models which account for rotation and effects attributed
to magnetic fields have been presented by Eggenberger, Maeder \&
Meynet (2006). Spruit (2002) and Braithwaite \& Spruit (2006) have
discussed the possibility of growing an azimuthal magnetic field via
stretching by differential rotation, followed by an instability, in
stably stratified radiative zones. Finally, Menou, Balbus \& Spruit
(2004) have generalized the diffusive stability analysis of GSF to the
case when a weak magnetic field is present in the radiative
zone. These diffusive rotational instabilities and their potential
role for the early Sun's radiative zone are our main subjects of
interest here.

The usual difficulty one faces when invoking fluid instabilities in
stellar radiative zones is that the strong radial thermal
stratification present stabilizes the fluid against a wide variety of
(adiabatic) perturbations. This is the case for rotational
instabilities in particular, even if strong radial differential
rotation (across spherical shells)\footnote{Any radiative zone with
negative differential rotation within a spherical shell ($\dd \ln
\Omega / \dd \theta < 0$) is subject to the standard
magneto-rotational instability because perturbations tangential to the
spherical shell are blind to the radial stratification, as shown by
Balbus \& Hawley (1994). Here, we are focusing exclusively on the more
challenging, ``orthogonal'' situation where negative differential
rotation exists across spherical shells ($\dd \ln \Omega / \dd \ln r <
0$; see Menou et al. 2004 for a discussion).} is present. For
instance, one verifies readily that, for any reasonable amount of
imposed radial differential rotation, the current Sun's radiative zone
would be stable according to both the general Solberg-H\o iland
adiabatic criteria (e.g. Tassoul 1978) and their generalization to
weakly-magnetized fluids (Balbus 1995).

A way around this difficulty has been proposed by Goldreich \&
Schubert (1967) and Fricke (1968), who have shown that rotational
instabilities can still exist, provided perturbed fluid elements are
allowed to exchange heat with their environment much faster than they
exchange momentum. For such (diabatic) perturbations, the stabilizing
role of thermal stratification is effectively neutralized when the
perturbed fluid element reaches thermal equilibrium with its
environment while its original momentum remains largely unchanged. In
the solar interior, radiative heat diffusion is indeed orders of
magnitude faster than viscous diffusion of momentum, thus motivating
the double-diffusive analysis of Goldreich \& Schubert and Fricke
(hereafter GSF altogether). The basic double-diffusive mechanism
invoked in these instabilities is related to that of ``salt-finger''
instabilities operating in the Earth's oceans, except that the
destabilizing salinity stratification is replaced here by a
destabilizing profile of angular momentum.

In Menou et al. (2004), we studied numerically the stability of the
current Sun's radiative zone to diffusive magneto-rotational
instabilities under a variety of imposed rates of radial differential
rotation. It was found that the Sun is relatively stable, unless a
fairly substantial rate of negative differential rotation is
imposed. However, as we alluded before, this is probably not the right
question to ask because angular momentum transport in the Sun's
radiative zone must have occurred in the past, when it was rotating
faster. One would generally expect this faster rotation to be
favorable to rotational instabilities. Therefore, we reconsider the
stability of the radiative zone of the (faster rotating, early) Sun
for various plausible rates of radial differential rotation. We do
find that, if the early Sun was rotating at least a few times faster
than it currently is, it was indeed much more prone to diffusive
hydro- and magneto-rotational instabilities.

We describe our method of solution in \S\ref{sec:met}. In
\S\ref{sec:res}, we present the results of our detailed stability
survey for the early Sun's radiative zone. In \S\ref{sec:conc} we
summarize key results, discuss some of their consequences and possible
extensions of this work.

\section{Method} \label{sec:met}

Our analysis is entirely based on a local, linear axisymmetric
stability analysis of differential rotation in a weakly-magnetized,
stably-stratified fluid. We recall the dispersion relation obeyed by
axisymmetric modes in a general context in \S\ref{ssec:disp}. The
choice of various physical parameters of importance in the dispersion
relation for the early Sun is described in \S\ref{ssec:params}. We
then describe how we obtained numerical solutions of the dispersion
relation in \S\ref{ssec:numsol}. In \S\ref{ssec:visc}, we discuss the
issue of the efficiency of momentum transport, that we estimate (or
rather extrapolate) from fastest growing linear mode properties.  Our
basic analysis follows closely that presented in Menou et al. (2004),
where additional details on the method can be found.

\subsection{Dispersion Relation} \label{ssec:disp}

The dispersion relation was derived in cylindrical coordinates, $({R},
{\phi}, {Z})$, but we will later switch to spherical coordinates, $(r,
\theta, \phi)$, for convenience. Axisymmetric Eulerian perturbations
with a WKB space-time dependence $\exp [i(\bb{k\cdot r} - \omega t)]$,
where $\bb{k}= (k_R, 0, k_Z) = (k_r,k_\theta,0)$ are considered.
Starting from the set of MHD equations (continuity, momentum,
induction, energy) that include the effects of viscosity, resistivity
and heat conduction,

\begin{eqnarray}
& & {\dd\rho\over \dd t} + \del\bcdot (\rho \bb {v}) =  0,
\end{eqnarray}
\begin{eqnarray}\label{mom}
& & \rho {\dd\vv \over \dd t} + (\rho \vv\bcdot\del)\vv = -\del\left(
P + {B^2\over 8 \pi} \right)  - \rho \del \Phi \nonumber \\
& & + \left( {\B\over 4\pi}\bcdot \del\right)\B + \mu \left( \del^2 \vv + 
\frac{1}{3} \del(\del \bcdot \vv) \right),
\end{eqnarray}
\begin{eqnarray}
& & {\dd\bb{B}\over \dd t} = \bb {\nabla\times (v\times B)} - 
  \eta { \bb \nabla\times} \left( {\bb \nabla\times \bb{B}} \right),
\end{eqnarray}
\begin{eqnarray}
& & {1\over (\gamma -1)} {P} {d\ln P\rho^{-\gamma}\over dt} = \chi \del^2 T,
\end{eqnarray}

one derives,  under the Boussinesq approximation and ignoring self-gravity, the following fifth-order dispersion relation for local,
axisymmetric, multi-diffusive modes, 

\begin{eqnarray}  
\nonumber & & {\tilde\omega_{b+v}}^4 {\omega_{e}}\, {k^2\over k_Z^2} + {\tilde\omega_{b+v}}^2 {\omega_{b}}
\left[ {1\over \gamma \rho}\, \left({\cal D} P\right)\, {\cal D} \ln
P\rho^{-\gamma}\right]\\ 
& & +{\tilde\omega_{b}}^2 {\omega_{e}} \left[
{1\over R^3}\, {\cal D} (R^4\Omega^2) \right] - 4 \Omega^2 (\kva)^2 {\omega_{e}}=
0,\label{eq:disprel}
\end{eqnarray}

where
\begin{eqnarray}
& &\bb{v_A} = { \bb{B}/\sqrt{4\pi\rho}}, \qquad 
k^2 = k_R^2 + k_Z^2,\nonumber \\ 
&& {\tilde\omega_{b+v}}^2=\omega_b \omega_v -(\kva)^2,
\qquad {\tilde\omega_{b}}^2=\omega_b^2 -(\kva)^2, \nonumber\\
\nonumber & & \omega_b=\omega+i \eta k^2, \qquad  \omega_v=\omega+i \nu k^2,\\
 \qquad
& & \omega_e=\omega+ \frac{\gamma -1}{\gamma}{i  T\over  P} \chi k^2, 
\qquad {\cal D} \equiv \left( \frac{k_R}{k_Z}\frac{\dd}{\dd Z}
-\frac{\dd}{\dd R}\right). \nonumber
\end{eqnarray}

A developed form of the dispersion relation can be found in Menou et
al. (2004). The \BV frequency, $N$, which measures the magnitude of
thermal stratification, comes out of the first bracket term in
Eq.~(\ref{eq:disprel}). Similarly, the epicyclic frequency, which
measures the angular momentum ``stratification,'' comes out of the
second bracket term. The notation is standard and identical to that
adopted in Menou et al. (2004): $\bb{v}$ is the fluid velocity, $\rho$
is the mass density, $P$ is the pressure, $\Phi$ is the gravitational
potential, $T$ is the temperature, $\bb{B}$ is the magnetic field,
$\mu$ is the dynamic viscosity, $\nu = \mu / \rho$ is the kinematic
viscosity, $\eta$ is the resistivity, $\chi$ is the heat conductivity
and $\bb{v_A}$ is the Alfv\'en speed.

A value $\gamma = 5/3$ is adopted for the gas adiabatic index and a
perfect gas equation of state is assumed. The basic state rotation is
given by $\bb{\Omega} = (0, 0, \Omega(R,Z))$ along the
${Z}$--axis. The basic state magnetic field, whose geometry is
specified below, is assumed to be weak with respect to both rotation
and thermal pressure.

\subsection{Parameters for the Early Sun} \label{ssec:params}

For our early Sun study, we adopt conditions which correspond
approximately to the zero-age main sequence Sun. Environmental
parameters in the early Sun's radiative zone which are important for
our stability analysis include local values of the density and
temperature, from which we determine the various diffusion
coefficients, and the value of the \BV frequency, which determines the
magnitude of thermal stratification.

Tables~\ref{tab:one} and~\ref{tab:two} list the values of all the
relevant parameters adopted in our stability survey, at two different
radii in the radiative zone of the early Sun. Note that these radii
($r = 0.3$ and $0.7$) are expressed in units of the early Sun's
radius, which is taken to be $0.87$ times that of the current Sun
($R_\odot$). More generally, quantities with a solar symbol subscript
refer to values for the current Sun. For instance, we consider a
faster rotating early Sun, with an angular velocity $\Omega \geq
\Omega_\odot \simeq 2.7 \times 10^{-6}$~rad~s$^{-1}$. Values for the
density, $\rho$, and temperature, $T$, in the radiative zone of the
early Sun were estimated from Bahcall, Pinsonneault \& Basu (2001). A
standard opacity table was used to deduce the radiative opacity,
$\kappa$. Values for the kinematic viscosity, $\nu$, radiative
viscosity, $\nu_r$, resistivity, $\eta$ and radiative diffusivity,
$\xi_{rad}$, were determined from standard relations for a fully
ionized gas, as described in Menou et al. (2004). Finally, the value
of the \BV frequency, $N$, was estimated from the evolutionary models
of Demarque \& Guenther (1991).

A comparison between our Tables~\ref{tab:one} \&~\ref{tab:two} and the
corresponding Table~1 in Menou et al. (2004) reveals that conditions
in the radiative zones of the early Sun and the current Sun are not
very different from each other, from the point of view of our
analysis. The ratio $\Omega_\odot^2 / N^2$ differs by a factor of a
few in the two cases, and so does the Prandtl number, $\epsilon_\nu
\equiv \nu / \xi_{rad}$, and the Acheson number, $\epsilon_\eta \equiv
\eta/ \xi_{rad}$. If anything, conditions are somewhat less
double-diffusive in the early Sun (with larger values of
$\epsilon_\nu$ and $\epsilon_\eta$) than in the current Sun. As we
shall see, it is then mostly the possibility that the early Sun was
rotating several times faster than it currently is which makes it
particularly sensitive to diffusive rotational instabilities.

\subsection{Numerical solutions} \label{ssec:numsol}

The dispersion relation (Eq.~[\ref{eq:disprel}]) is of such complexity
that it is difficult to derive from it general necessary and
sufficient conditions for stability from it (Menou et al. 2004). In
addition, we are interested here in studying the properties of
unstable modes, not just the extent of stability domains. We thus
solve the dispersion relation numerically, with a set of parameters
appropriate for the early Sun (\S\ref{ssec:params}).  This is done
using the Laguerre algorithm described by Press et al. (1992) to solve
Eq.~(\ref{eq:disprel}) as a fifth-order complex polynomial for
$\sigma=-i \omega$, with real coefficients.

It is useful to rewrite both the rotational and thermal stratification
bracket terms appearing in the dispersion relation in spherical
coordinates $(r, \theta, \phi)$. The thermal stratification term then
becomes function of $N^2$, $\theta$ and the radial and angular
wavevectors, $k_r$ and $k_\theta$, for a spherically symmetric star.
The rotational stratification term, on the other hand, explicitly
depends on the amount of differential rotation within and across
spherical shells, that we express as $\dd \ln \Omega / \dd \theta$ and
$\dd \ln \Omega / \dd \ln r$, respectively. Here, we are interested in
the specific case corresponding to $\dd \ln \Omega / \dd \theta = 0$
(no differential rotation within a spherical shell) but various
degrees of negative differential rotation across spherical shells
($\dd \ln \Omega / \dd \ln r < 0$). This choice is dictated by the
fact that weakly-magnetized radiative zones with any $\dd \ln \Omega /
\dd \theta < 0$ are subject to the standard (adiabatic)
magneto-rotational instability (Balbus \& Hawley 1994). To explain the
current solid body rotation of the Sun's radiative zone, it is the
more challenging $\dd \ln \Omega / \dd \ln r < 0$ case which is also
most relevant, given the general expectation of faster rotating inner
regions (\S\ref{sec:intro}).

We search for unstable modes (which correspond to $\sigma$-roots with
positive real parts) at various locations on the sphere within the
radiative zone.  We consider values of the polar angle, $\theta$, in
the range $0 \to \pi/2$ (pole to equator). At each location, we
perform a search for the fastest growing unstable mode by varying the
wavevector components $k_r$ and $k_\theta$ independently in the range
$\pm [2 \pi / l_{\rm max}, 2 \pi / l_{\rm min} ]$, on a $400 \times
400$ grid in $k$-space. The first, large-scale limit, $l_{\rm max}$,
is chosen to be $\ltsim H \simeq \Re T/ g$, the pressure scale height
(where $\Re$ is the perfect gas constant and $g$ is the local
gravitational acceleration). This guarantees that the local and weak
field assumptions of our analysis remain valid. The second,
small-scale limit, $l_{\rm min}$, is chosen to be $ \gg \lambda$, the
mean free path, so that the analysis remains valid given our use of
fluid equations. The exact value adopted for $\lambda$ turns out to be
unimportant, as we find that all modes tend to be stabilized by
diffusion on such extremely small spatial scales anyway.

It is generally difficult to guarantee that one has successfully
identified, with a discrete search in $k$-space, the fastest growing
mode for a given set of environmental parameters in the dispersion
relation. However, we have performed extensive numerical convergence
tests (at twice and four times the resolution in each dimension) which
suggest that our results are indeed converged. At less than half the
standard resolution adopted, on the other hand, deviations from our
converged results start becoming significant.  We find that the
identification of fastest growing modes is facilitated (in terms of
numerical resolution) as conditions become increasingly more unstable
(e.g., an increased rate of differential rotation).  Additional tests
showing that our coding of the dispersion relation behaves as expected
in various idealized limits have been described in Menou et al. (2004;
see also Menou 2004). As mentioned in these studies, it is possible to
verify explicitly the diffusive nature of unstable modes by imposing
very small changes in the values of the diffusion coefficients and
being able to witness small, but discernible effects on the mode
properties. This is in contrast with the lack of such dependence in
the case of adiabatic modes (which emerge, e.g., when $\dd \ln \Omega
/ \dd \theta < 0$).

\subsection{Efficiency of Transport: ``Turbulent Viscosity''}  \label{ssec:visc}

It is unclear whether one can determine, from a linear analysis only,
the efficiency of transport that would result from non-linear growth
and turbulence. We will adopt an often-used prescription here, by
assuming that the efficiency of transport can be crudely estimated
from the growth rate and the wavelength of the fastest-growing
unstable linear mode. There is no general justification for this
simple prescription, which may be off even at the order of magnitude
level. Our goal here, however, is not to estimate too accurately the
efficiency of momentum transport between shells, but rather to
determine whether, qualitatively, diffusive rotational instabilities
could be responsible for angular momentum transport at the level
required to affect globally a pattern of differential rotation in the
Sun's radiative zone.

It should be noted that the non-linear numerical simulations of
Korycansky (1991) lend support to the simple prescription adopted
here. This author found, for an equatorial setup (i.e. aligned thermal
and angular momentum stratifications), that the non-linear behavior of
hydrodynamical double-diffusive (GSF) modes is well described by a
combination of linear growth rate and radial wavelength of the
instability. Assuming this property carries over to magnetized modes
and to other geometries (i.e. away from the stellar equator), we will
be estimating the efficiency of angular momentum transport across
spherical shells with the quantity $\sigma_{\rm max}/k^2_{r,\,{\rm
max}}$ (dimension of a viscosity), where $\sigma_{\rm max}$ is the
growth rate of the fastest-growing mode and $k_{r,\,{\rm max}}$ is the
spherical-radial wavevector component of that same mode. We will refer
to this quantity as ``turbulent viscosity'' in what follows.  Clearly,
using linear theory to predict a non-linear outcome is a daring
extrapolation, so one should treat our results on transport
efficiencies with extreme caution.

With this caveat in mind, a global measure of the efficiency of
angular momentum transport required in the Sun's radiative zone is
obtained from the minimum turbulent viscosity value needed to affect
differential rotation on a characteristic timescale, $\tau$. Choosing
a billion years as a reference turbulent viscous time, one requires a
turbulent viscosity
\begin{equation} \label{eq:tvis2}
\nu_{\rm turb} > \frac{R_\odot^2}{\tau = 10^9~{\rm years}} \sim 1.5
\times 10^5~{\rm cm^2~s^{-1}}.
\end{equation}
This minimum value of $\nu_{\rm turb}$ will be useful in comparison to
turbulent viscosity values estimated from our stability survey. Notice
that this value is indeed several orders of magnitude larger than the
values of the microscopic viscosities $\nu$ and $\nu_r$ listed in
Table~\ref{tab:one}.

\section{Results} \label{sec:res}

In our analysis, the two main parameters determining the stability of
the early Sun's radiative zone are its angular velocity $\Omega$
($\geq \Omega_\odot$) and the rate of negative differential rotation
across spherical shells ($\dd \ln \Omega / \dd \ln r < 0$) assumed to
be present. While one can safely assume that the Sun was not rotating
much faster than $30$-$50$ times its current rate in the past, it is
much more difficult to estimate what level of radial differential
rotation may have been present. Noting that the Keplerian shear rate,
which is strong even by astronomical standards, would correspond to
$\dd \ln \Omega / \dd \ln r = -3/2$ along the equator, we
systematically survey the parameter space for values of $\dd \ln
\Omega / \dd \ln r$ ranging from $0$ to $-2.5$. Independently, we vary
the value of the local angular velocity $\Omega$ from $ \Omega_\odot
$to $31 \, \Omega_\odot$.

Even with these choices, the number of free parameters in our
stability analysis remains large. For simplicity, we focus in
\S\ref{ssec:hyd} and \S\ref{ssec:mhd} on special cases, which turn out
to be both illustrative and representative.  They correspond to a
specific choice of location/geometry in the early Sun ($r=0.7$,
$\theta = \pi/4$) and to specific magnetic field conditions ($B=0$ or
$B_\theta =4$~G).  We then discuss in \S\ref{ssec:bdep} and
\S\ref{ssec:odep} additional parameter dependencies as they emerged
from our extended survey.

\subsection{Fastest Growing GSF Modes} \label{ssec:hyd}

The dispersion relation (Eq.~[\ref{eq:disprel}]) can be solved in the
hydrodynamic limit ($v_A=\eta=0$), in which case it reduces exactly to
the dispersion relation of Goldreich \& Schubert (1967). In this
limit, we can thus examine the stability properties of the early Sun's
radiative zone with respect to GSF modes.

Figure~\ref{fig:one} shows the growth rate, $\sigma$ (in units of the
angular velocity $\Omega$; left panel), and the corresponding
``turbulent viscosity,'' $\sigma / k_r^2$ (in units of
cm$^2$~s$^{-1}$; right panel), of the fastest growing GSF mode, under
early Sun conditions at $r=0.7$ and $\theta = \pi/4$. In both panels,
color contours of $\sigma$ and $\sigma / k_r^2$ as a function of the
assumed angular velocity, $\Omega / \Omega_\odot $ (horizontal axis),
and rate of negative differential rotation, $\dd \ln \Omega / \dd \ln
r$ (vertical axis), are shown.

In an early Sun rotating at the same rate as the current Sun, GSF
instabilities appear only for relatively strong rates of radial
differential rotation. As the rotation rate is increased to a few
times the current one, however, GSF instabilities are much more easily
triggered, and for values of $\Omega / \Omega_\odot \gsim 5$--$10$,
any radial differential rotation stronger than $\dd \ln \Omega / \dd
\ln r \lsim -0.25$ is GSF--unstable. Notice how the growth rates, in
units of $\Omega$, become of order unity and independent of $\Omega /
\Omega_\odot$ for values of this parameter $\gsim 10$--$15$.

It is remarkable that the current rotation rate of the Sun corresponds
approximately to the stability limit for GSF modes (as shown by the
sharp drop in contours as $\Omega / \Omega_\odot \to 1$ in
Fig.~\ref{fig:one}). There is nothing that guarantees this a priori in
our stability analysis. As we shall see later, this is related to a
close equality between dynamical and diffusive timescales for current
Sun conditions, a coincidence which was already commented upon by,
e.g., Acheson (1978). We will revisit this issue in \S\ref{sec:conc}.

Despite their prominence in the faster-rotating early Sun, GSF modes
may not have been able to provide efficient enough angular momentum
transport to modify the existing differential rotation rate. As shown
in the right panel of Fig.~\ref{fig:one}, turbulent viscosities
associated with fastest growing GSF modes tend to be small: $\nu_{\rm
turb} \sim 10^{1-4}$~cm$^2$~s$^{-1}$. According to
Eq.~(\ref{eq:tvis2}), this is not sufficient to provide much transport
over the $5$ billion years lifetime of the Sun. Another limitation of
GSF modes, as we have already mentioned, is that they are not expected
to bring the fluid to a state of solid body rotation (at marginal
stability). This is in contrast with the MHD versions of these
diffusive modes, which have the potential to lead to solid body
rotation (at marginal stability; Menou et al. 2004).

\subsection{Fastest Growing MHD Modes}\label{ssec:mhd}

Figure~\ref{fig:two} shows growth rate and turbulent viscosity values
obtained for the fastest growing MHD modes in the presence of a purely
polar magnetic field of strength $B_\theta =4$~G, for the same early
Sun conditions at $r=0.7$ and $\theta = \pi/4$ as before. While the
contour plots are broadly consistent with those in Fig.~\ref{fig:one},
important quantitative differences emerge. The MHD nature of the
fastest growing modes under these conditions can be verified by
changing slightly the magnetic field strength and being able to
witness small, but discernible effects on the mode properties (see
also our discussion of the effects of magnetic field strength in
\S\ref{ssec:bdep}).

Fastest growing MHD modes which have very slow growth rates have been
voluntarily truncated from the $\sigma$ contour plot (left panel,
Fig.~\ref{fig:two}), for values of $\sigma/\Omega < 0.01$. This
greatly facilitates the identification of regions which are only
marginally unstable, by comparison with the corresponding contours of
turbulent viscosity in the right panel of Fig.~\ref{fig:two} (which
include all fastest growing modes, no matter how slowly they are
growing). It is clear from this comparison of the two panels in
Fig.~\ref{fig:two} that substantial regions of the parameter space
surveyed for MHD modes correspond to fastest growing modes which are
in fact growing quite slowly. Contrary to GSF modes
(Fig.~\ref{fig:one}), only for the largest values of $\Omega /
\Omega_\odot$ and $\dd \ln \Omega / \dd \ln r$ do the growth rates,
$\sigma/ \Omega$, approach unity for MHD modes. Note also that
unstable MHD modes only appear for values of $\Omega / \Omega_\odot
\gsim 3$--$4$. This leads us to conclude that, in the presence of a
magnetic field, it is more difficult to trigger diffusive rotational
instabilities (for the conditions in the early Sun's radiative zone).

This difficulty comes with an attractive feature, however, from the
point of view of transport. As the right panel of Fig.~\ref{fig:two}
shows, fastest growing MHD modes tend to have larger associated
turbulent viscosity values than their hydrodynamic GSF counterparts,
by orders of magnitude (compare right panel of
Fig.~\ref{fig:one}). Even though fastest growing MHD modes tend to
grow more slowly than GSF modes, they do so on larger scales than the
corresponding fastest growing GSF modes, which results in an
effectively larger turbulent viscosity estimate for the MHD case (at
least according to our extrapolation from linear theory).

This difference between MHD and hydro (GSF) modes can be understood
simply as being a consequence of the destabilizing role played by
magnetic tension in MHD modes. Our diffusive MHD modes are
intrinsically magneto-rotational in nature. The role of magnetic
tension for the magneto-rotational instability has been documented
extensively (e.g. Balbus \& Hawley 1998). It is well known that, for a
given magnetic field strength and orientation, magnetic tension picks
a specific ``most unstable'' scale, below which it ultimately acts as
a stabilizing agent (i.e. like a slow MHD wave) and above which it
becomes very inefficient at destabilizing the flow. Provided that this
best scale for destabilizing magnetic tension is much larger than the
diffusive scales in the problem, one would expect rather slowly
growing MHD modes (since diffusion, slower on larger scales, is still
necessary for any instability), operating on larger scales than their
hydrodynamical (GSF) counterparts. This interpretation is consistent
with all our survey results for fastest growing MHD and GSF modes.

\subsection{Dependence on the Magnetic Field Strength} \label{ssec:bdep}

One would also expect the hydrodynamic nature of the fastest growing
modes to be recovered for small enough values of the magnetic field
strength, when the ``most unstable'' scale picked by magnetic tension
becomes comparable to the typical diffusive scales associated with GSF
modes. We indeed observe such a hydro-to-MHD transition in our
parameter space survey.

\begin{figure}
\plottwo{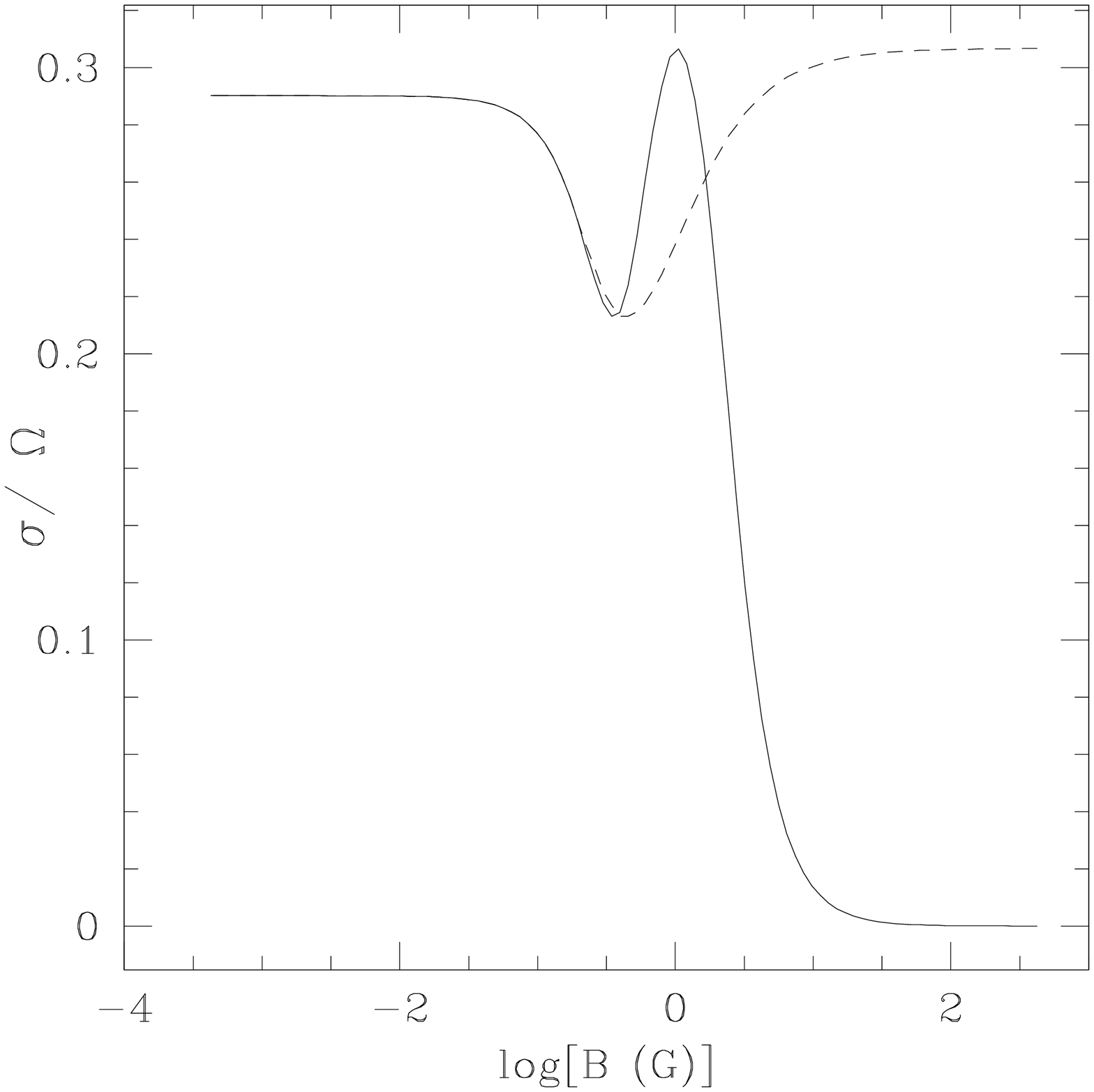}{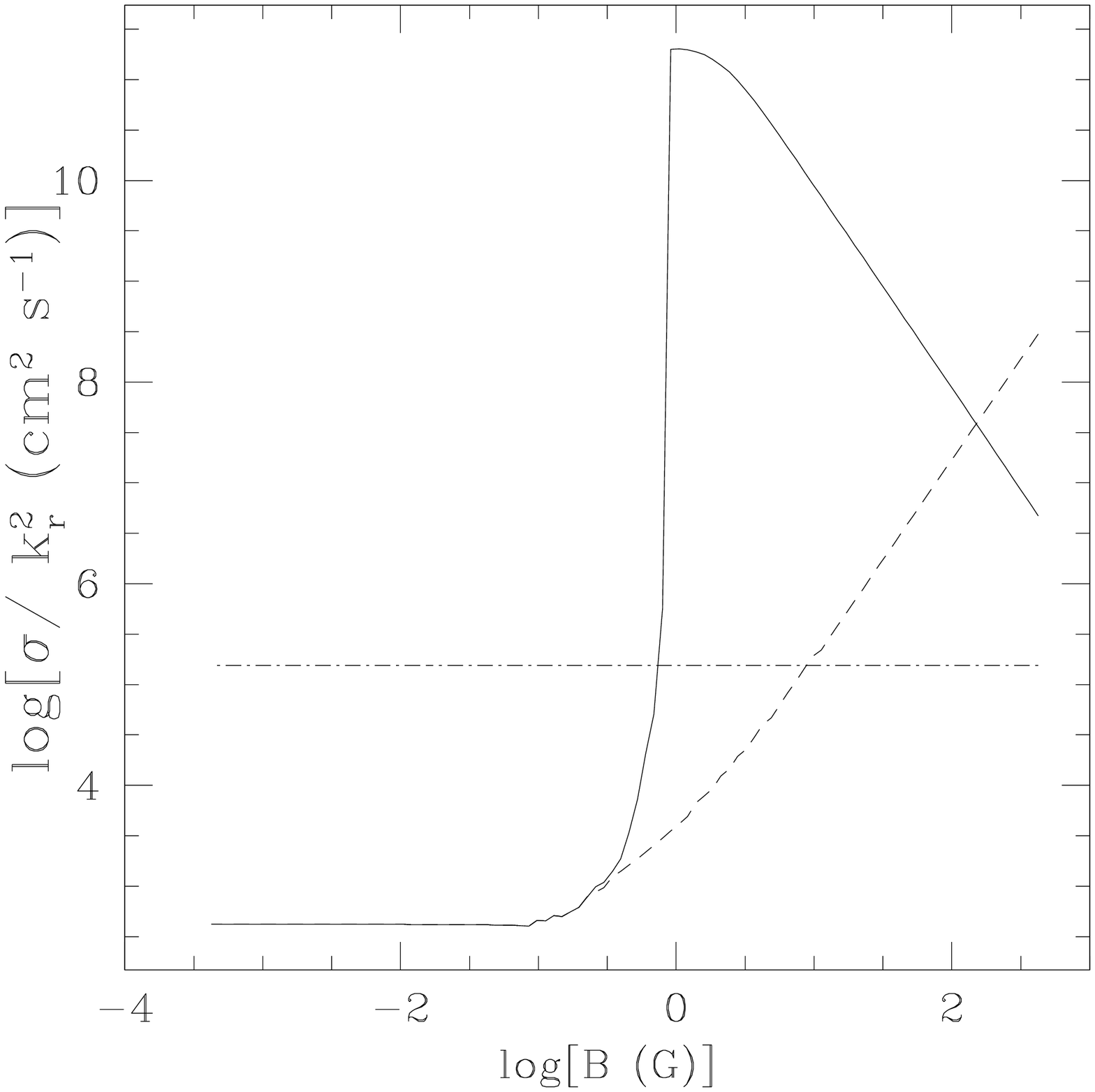}
\caption[]{\label{fig:three} Growth rate (left, $\sigma / \Omega$) and
``turbulent viscosity'' (right, $\sigma / k_r^2$) of the fastest
growing mode, as a function of the strength and geometry of the
magnetic field, $B$. Solid lines are for a purely polar field
($B_\theta$ only) and dashed lines for a purely radial field ($B_r$
only). A rotation rate $\Omega =20 \, \Omega_{\rm sun}$ and a rate of
differential rotation across spherical shells $d\ln\Omega / d\ln r
=-1$ were assumed. The analysis is performed at a specific location in
the early Sun's radiative zone, corresponding to a polar angle $\theta
= \pi /4$ and a spherical radius $r = 0.7$ (in units of the early Sun
radius). Hydrodynamical GSF modes are recovered at low enough values
of the magnetic field strength.  The horizontal dash-dotted line in
the right panel indicates the angular momentum transport efficiency
required for a global transport timescale in the Sun $\sim 10^9$~yr.}
\end{figure}

Figure~\ref{fig:three} shows how the properties of the fastest growing
mode are affected by changes in the assumed strength of the magnetic
field. For ease of comparison, the same early Sun conditions as
before, at $r=0.7$ and $\theta = \pi/4$, have been adopted here as
well. Furthermore, the local rates of rotation and radial differential
rotation have been fixed to $\Omega / \Omega_\odot =20$ and $\dd \ln
\Omega / \dd \ln r = -1$. We are thus effectively looking at a single
point in the contour plots of Fig.~\ref{fig:one} and~\ref{fig:two} and
studying how the transition from the hydrodynamic to the MHD regime
proceeds as the magnetic field strength is
varied. Figure~\ref{fig:three} shows how this transition happens for a
purely radial field ($B_r = 10^{-3.5}$--$10^{+2.5}$~G; dashed lines)
and a purely polar field ($B_\theta =10^{-3.5}$--$10^{+2.5}$~G; solid
lines). The effects of the hydro-to-MHD transition are monitored in
terms of the same normalized growth rate, $\sigma / \Omega$ (left
panel), and turbulent viscosity, $\sigma / k_r^2$ (right panel), for
the fastest growing mode as before.

For low enough values of the magnetic field ($B_r$, $B_\theta \ll
0.1$~G), the fastest growing mode properties become independent of the
magnetic field strength and one verifies that the results for purely
hydrodynamic GSF modes are indeed recovered. As the value of $B$ is
increased beyond $0.1$~G, the first effect is a reduction in the
growth rate of the most unstable mode, followed quickly by an increase
for $B \gsim 0.5$~G. From then on, the effect of increasing $B$
depends on the field geometry. A larger field tends to lead to faster
growing MHD modes and to larger values of the turbulent viscosity (as
measured by $\sigma / k_r^2$). Although this trend ends and reverses
in the case of the strictly polar field (solid line), for $B_\theta
\gsim 1$~G, we shall see below, when we discuss Fig.~\ref{fig:four},
that this change of behavior is mostly an artifact of our method, which
signals the breakdown of the local theory adopted.

A key feature of the transition from hydro to MHD diffusive modes is
the increase in the turbulent viscosity value ($\sigma / k_r^2$). The
horizontal dash-dotted line in the right panel of Fig.~\ref{fig:three}
indicates the minimum value of $\nu_{\rm turb}$ required for global
transport in the Sun's radiative zone on a timescale $\sim 10^9$~yr
(Eq.~[\ref{eq:tvis2}]). While the turbulent viscosity that we estimate
for hydro GSF modes seems insufficient to have had an effect on the
internal rotation profile of the early Sun's radiative zone, diffusive
modes in the MHD regime ($B \gg 0.1$~G) are apparently capable of
transporting angular momentum efficiently enough.

Figure~\ref{fig:four} illustrates two additional properties of these
fastest growing unstable modes as they make the transition from hydro
to MHD (same notation as in Fig.~\ref{fig:three}). The ratio of their
radial to polar wavenumbers is shown in the left panel, while the
ratio of their radial wavelength to the local pressure scale height is
shown in the right panel. Fastest growing hydro (GSF) modes tend to
have comparable radial and polar wavenumbers, while MHD modes have
predominantly polar wavenumbers (and thus polar wavevectors, which
correspond to predominantly radial displacements). In addition, while
hydro GSF modes are confined to rather small spatial scales relative
to the local pressure scale height ($k_{\rm H}/k_r \ll 1$), fastest
growing MHD modes have increasingly larger radial wavelengths as the
magnetic field strength is increased, as expected if the spatial scale
favored by the destabilizing magnetic tension is forced to larger
scales by the stronger magnetic field.

As explained in \S\ref{ssec:numsol}, we have voluntarily limited our
search for fastest growing unstable modes to wavelengths smaller than
the local pressure scale height. This limit is manifestly reached in
the case of large enough polar field strengths (solid line in the
right panel of Fig.~\ref{fig:four}) and, retrospectively, it is clear
that this limit being reached affects all the other properties shown
in Figs.~\ref{fig:three} and~~\ref{fig:four} for this specific field
geometry (when $B_\theta \gsim 1$~G). Although this may signal the
breakdown of the local stability theory used, it is clear from the
trends in Figs.~\ref{fig:three} and~~\ref{fig:four} that
extrapolations beyond the local limit would tend to reinforce our main
conclusions: fastest growing diffusive MHD modes have larger
wavelengths and appear to be more efficient at transporting angular
momentum across spherical shells than their hydrodynamical (GSF)
counterparts.

\subsection{Other Parameter Dependencies} \label{ssec:odep}

We have verified that the trends illustrated in
Figs.~\ref{fig:one}--\ref{fig:four} remain generally valid throughout
the parameter space surveyed. We have varied the geometry of the
stability problem by picking values of $\theta$ from $0$ to
$\pi/2$. While rotational instabilities tend to be somewhat favored
(or stronger) at $\theta = \pi/2$ (equator), the difference with the
$\theta = \pi/4$ is small. As $\theta \to 0$ (pole), rotational
instabilities relying on radial differential rotation are disfavored,
as one expects from simple geometrical considerations. This points to
a general asymmetry in transport between equatorial and polar regions,
but we have not explored this trend more systematically.

We have also explored stability at a different location, $r=0.3$, in
the early Sun's radiative zone. As expected from the scalings in
Tables~\ref{tab:one} and~\ref{tab:two}, conditions are less
double-diffusive (and thus less prone to instability) at this radius
than at $r=0.7$. For $r=0.3$, we find growth rate and turbulent
viscosity contours which are roughly consistent with those shown in
Figs.~\ref{fig:one} and~\ref{fig:two}, except that they are all
shifted to larger values of $\Omega/\Omega_\odot$, by a factor
$4$--$5$. For instance, GSF modes do not appear in our survey at
$r=0.3$ before values of $\Omega \gsim 5 \Omega_\odot$ (vs. $\Omega
\simeq \Omega_\odot$ in Fig.~\ref{fig:one}). Similarly, for the same
MHD conditions as in Fig.~\ref{fig:two}, values of $\Omega
/\Omega_\odot \gsim 15$--$20$ are required to trigger instability at a
rate of differential rotation $\dd \ln \Omega / \dd \ln r =-1$, and
values of $\Omega /\Omega_\odot \gsim 30$ are required for instability
if $\dd \ln \Omega / \dd \ln r =-0.5$. In all aspects except these
changes in the location of the stability domains, our results for
$r=0.3$ are comparable to those at $r=0.7$.

\section{Discussion and Conclusion} \label{sec:conc}

The main conclusions of our stability survey for the early Sun's
radiative zone can be summarized as follows. Provided that the early
Sun was rotating at least a few times faster than it currently is, we
find that it was easily subject to diffusive hydro- and
magneto-rotational instabilities. That is, rather weak levels of
negative radial differential rotation (across spherical shells) were
sufficient to trigger these linear instabilities. This result
complements the one stating that any level of negative differential
rotation within a spherical shell ($\dd \ln \Omega / \dd \theta < 0$)
triggers the standard (adiabatic) magneto-rotational instability
within that shell (Balbus \& Hawley 1994). An inspection of our survey
results suggests that diffusive magneto-rotational instabilities may
be more efficient at transporting angular momentum than their
hydrodynamical (GSF) counterparts because fastest growing unstable
modes typically have larger wavelengths in the magnetized case. A
simple interpretation of this trend in terms of the scale favored by a
destabilizing magnetic tension was offered. This leads us to conclude
that diffusive magneto-rotational instabilities are attractive
candidates to explain the angular momentum transport (and possibly the
elemental mixing) that must have occurred in the Sun's radiative zone.

Additional work will be required to establish these promising trends
on firmer grounds. Indeed, a number of questions raised by our
investigation remain unanswered at this point. Chief among them is the
issue of the primordial strength (and to some extent geometry) of the
magnetic field present in the Sun's radiative zone.  Our study of the
hydro-to-MHD transition shows that if the magnetic field initially
present in the Sun's radiative zone was too weak, the fluid would have
behaved largely as if it were unmagnetized, as far as rotational
instabilities are concerned.\footnote{We note that there is a
well-known pathology if one attempts to recover the stability
conditions of an unmagnetized fluid by taking the zero-field limit of
the corresponding ideal MHD stability conditions (e.g. Balbus
1995). Our results are interesting in this respect, as they show how
the proper unmagnetized limit can be recovered at small, but finite,
field values, when diffusive effects are properly accounted for.}  The
minimum magnetic field values needed for the fluid stability to be
determined by MHD processes are not large ($B \gsim 0.1$ G) and they
do not violate current upper limits on the field strength in the Sun's
radiative zone (e.g. Friedland \& Gruzinov 2004 and references
therein). However, if the Sun's radiative zone was initially seeded
with only a very weak magnetic field, the growth of GSF instabilities
would be favored and the resulting transport may then be
weak. Diffusive magneto-rotational instabilities could eventually play
a role if these GSF instabilities were able to grow the background
seed field to a large enough value for MHD modes to take
over. Although this hypothetical scenario shares some similarities
with the kinematic dynamo problem, the likelihood of its success
remains very unclear at this point.

\begin{figure}
\plottwo{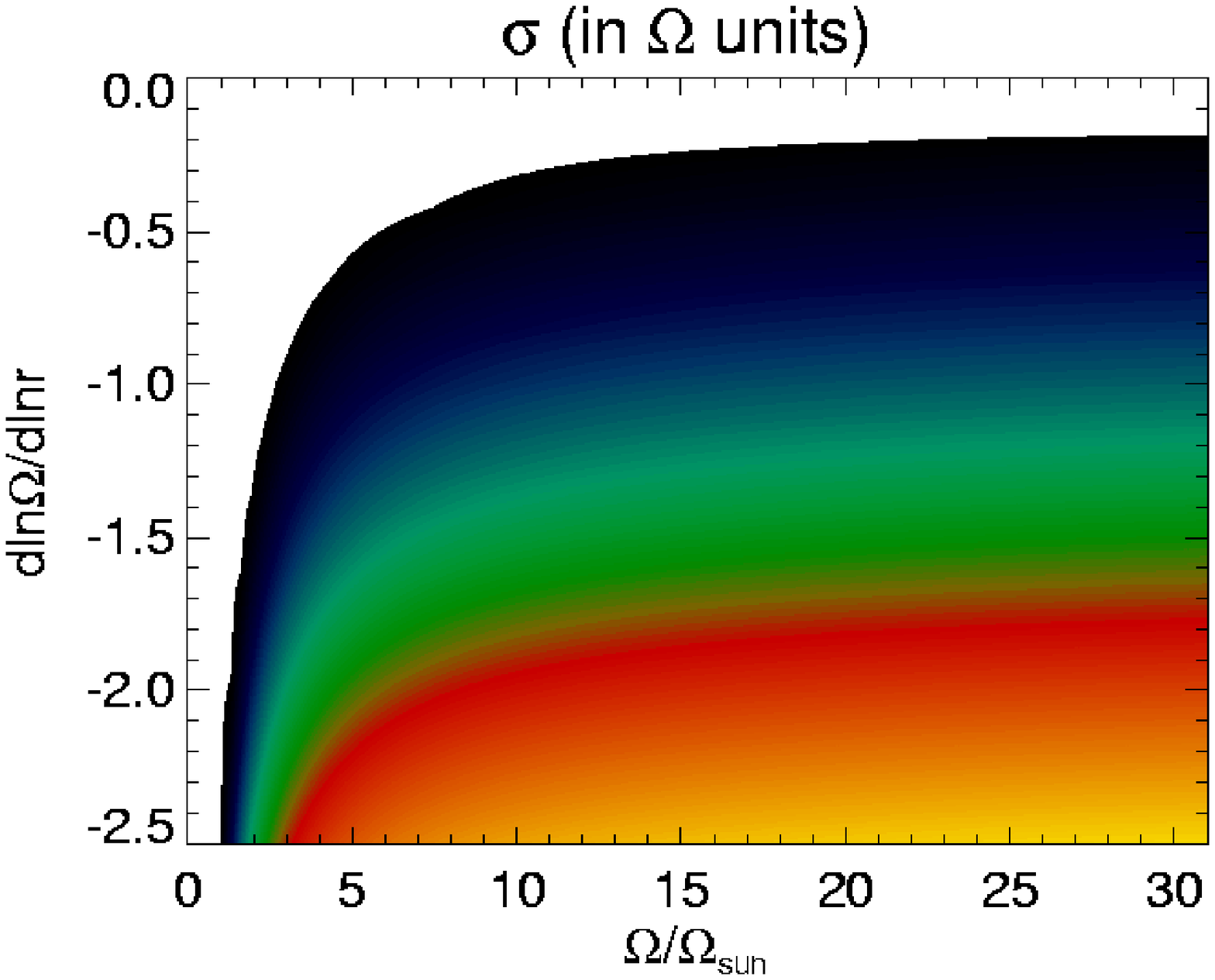}{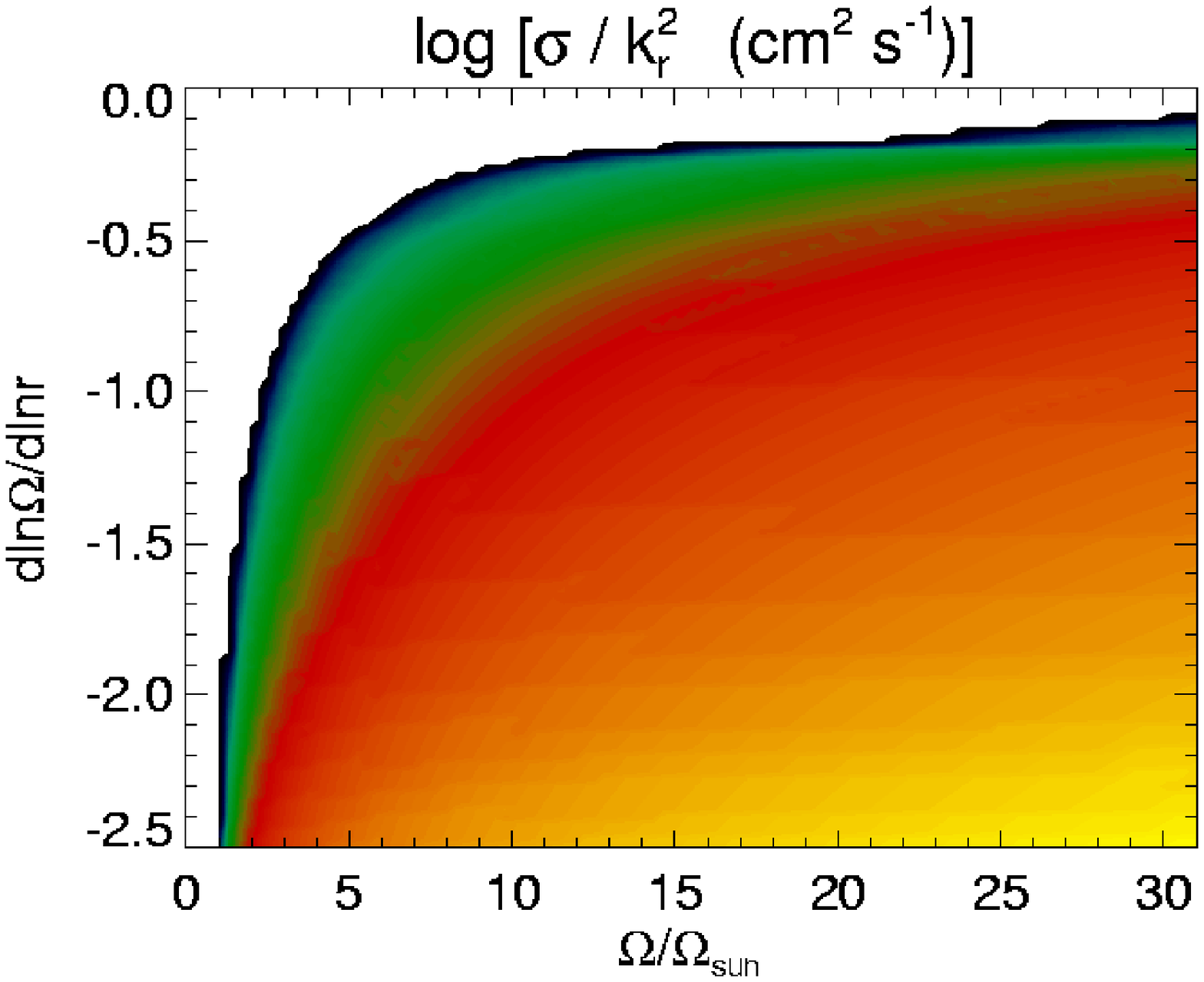}
\caption[]{\label{fig:one} Growth rate (left, $\sigma/\Omega$) and
``turbulent viscosity'' (right, $\sigma/k_r^2$) of fastest growing
hydrodynamical (``GSF'') modes, as a function of the rotation rate
($\Omega/ \Omega_{\rm sun}$) and the level of differential rotation
across spherical shells ($d\ln\Omega / d\ln r$). The stability survey
is performed at a specific location in the early Sun's radiative zone,
corresponding to a polar angle $\theta = \pi /4$ and a spherical
radius $r = 0.7$ (in units of the early Sun radius, see
Table~\ref{tab:one}).}
\end{figure}

\begin{figure}
\plottwo{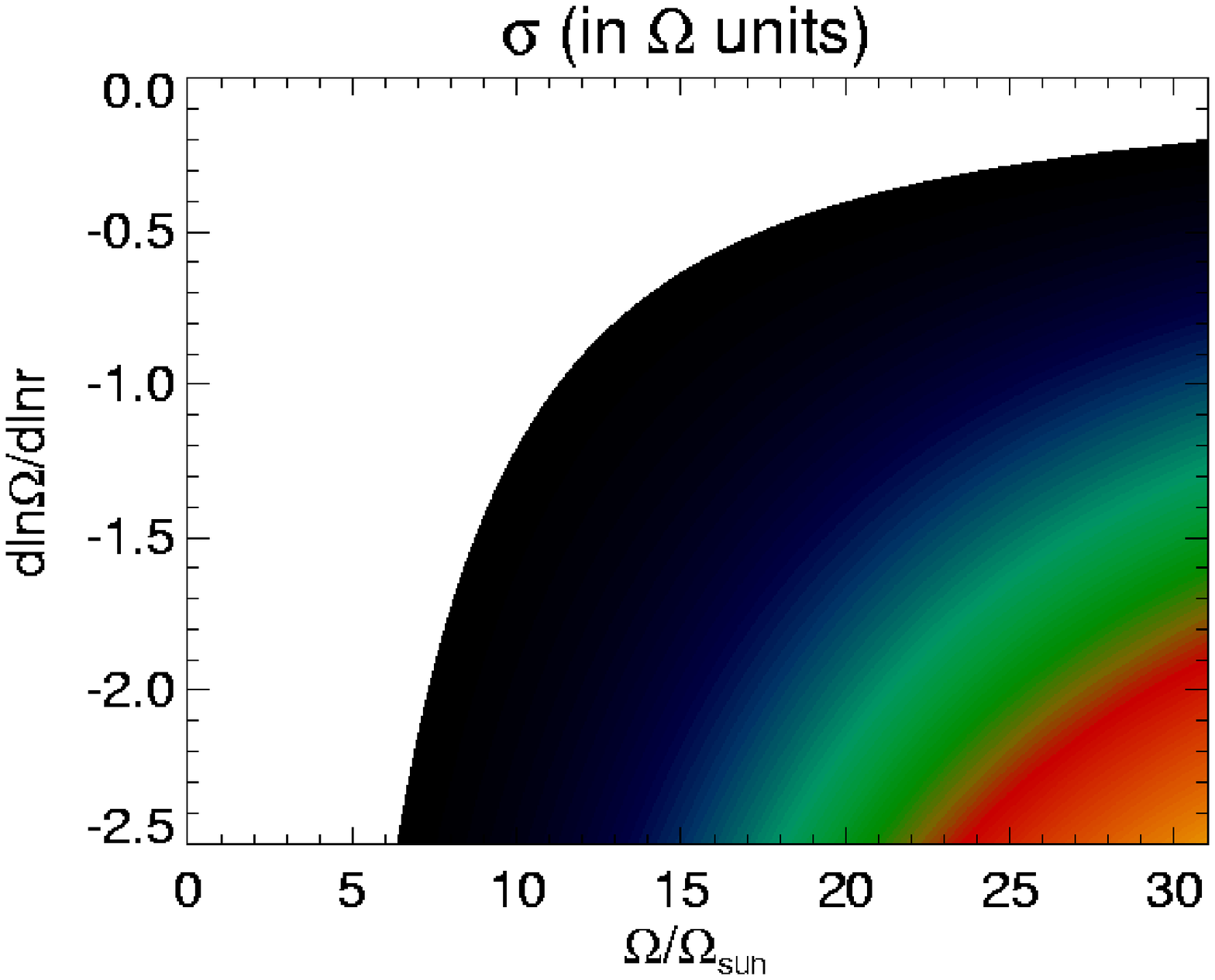}{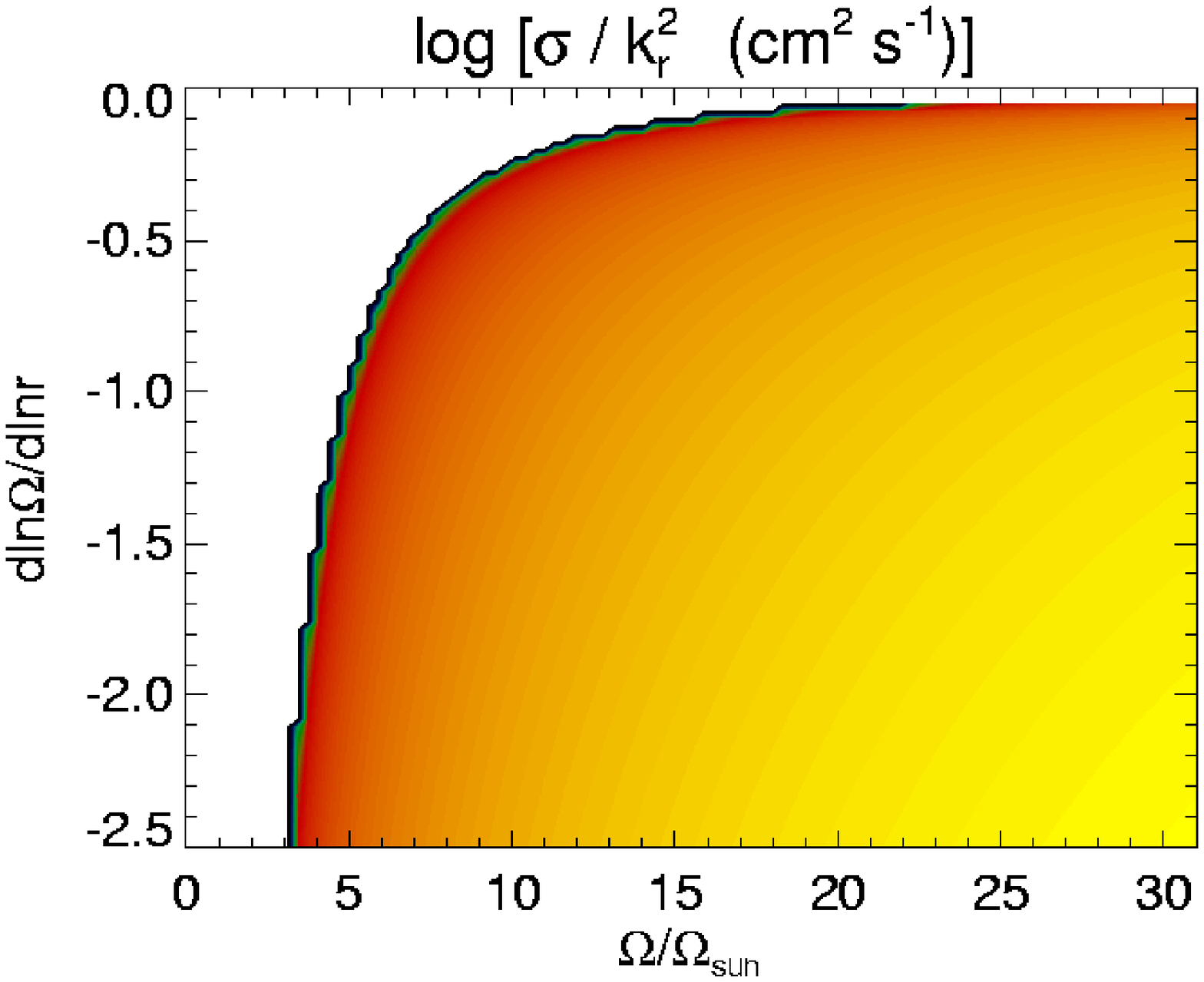}
\caption[]{\label{fig:two} Growth rate (left, $\sigma / \Omega$) and
``turbulent viscosity'' (right, $\sigma / k_r^2$) of fastest growing
MHD modes, as a function of the rotation rate ($\Omega/ \Omega_{\rm
sun}$) and the level of differential rotation across spherical shells
($d\ln\Omega / d\ln r$). The stability survey is performed at a
specific location in the early Sun's radiative zone, corresponding to
a polar angle $\theta = \pi /4$ and a spherical radius $r = 0.7$ (in
units of the early Sun radius), and for a specific strength and
geometry of the local magnetic field (polar field $B_\theta =
4$~G). Note that the linear color scale in the left panel truncates
modes growing at a rate $\sigma$ slower than $0.01 \, \Omega$.}
\end{figure}

In addition to this dependence on seed field conditions, the role of
diffusive rotational instabilities is likely to be affected by the
chemical structure and evolution of the radiative zone in which they
operate. By analogy with the result of Goldreich \& Schubert (1967) on
the strongly stabilizing role of composition gradients, one would
generally expect diffusive rotational instabilities to be quenched
when substantial composition gradients start developing in a radiative
zone. This may limit to very early times only their role for the
innermost radiative core of the Sun and it points, more generally, to
the possibility of a complex, coupled chemical-rotational evolution
(since turbulent mixing could affect the magnitude of the stabilizing
composition gradients being established). In regions where composition
gradients end up being significant (like the current Sun inner
radiative core), it appears likely that the dynamo and transport
process advocated by Spruit (2002) and simulated by Braithwaite \&
Spruit (2006) would be dominant and free to operate (see also Heger,
Woosley \& Spruit 2005 for an application to massive star
evolution). In regions with negligibly small composition gradients,
however, diffusive modes, growing exponentially with time constants
approaching the local value of the angular velocity
(Figs.~\ref{fig:one} and~\ref{fig:two}), are likely to dominate over a
linear growth by field stretching, which is invoked in the initial
phase of the scenario proposed by Spruit (2002; see also Braithwaite
\& Spruit 2006).

Let us conclude by mentioning that the locations of stability domains,
as found in our survey, are broadly consistent with a simple argument
involving a balance between local diffusive and dynamical timescales.
Acheson (1978) already emphasized that instability to double-diffusive
GSF modes should be approximately determined by the criterion
\begin{equation}
\frac{\nu}{\xi} \lsim - \frac{\kappa^2}{N^2},
\end{equation}
where $\kappa$ is the epicyclic frequency (which must be imaginary for
instability). This simply states that the stabilizing role of thermal
stratification ($N^2$) must be neutralized by a strong enough heat
diffusivity ($\xi$), but that at the same time the destabilizing role
of the angular momentum profile ($\kappa^2 < 0$) must not be
neutralized by momentum diffusion (i.e. viscosity, $\nu$). This simple
criterion appears to be approximately satisfied in our stability
survey. Similarly, our survey results indicate that instability to
double-diffusive MHD modes is roughly determined by a similar
criterion,
\begin{equation}
\frac{\eta}{\xi} \lsim -\frac{\dd \Omega^2 / \dd \ln r}{N^2},
\end{equation}
where $\eta$ is the resistivity. This explains why unstable MHD modes
only appear for values of $\Omega / \Omega_\odot$ which are a few
times larger than for GSF modes, in approximate proportion with the
ratio $\sqrt{\eta/\nu}$. It would be interesting to apply these simple
criteria to the radiative zones of other stars than the Sun, at
different stages of their evolution, to determine more generally how
relevant diffusive rotational instabilities may be for the evolution
of rotating stars.

\clearpage
\begin{table}[htdp]
\caption{Conditions in the Early Sun's Radiative Zone}
{\footnotesize
\begin{center}
\begin{tabular}{cccccccc}
\hline
\hline
\\
Radius & $\rho$&$T$&$\kappa$&$\nu$&$\nu_r$&$\eta$&$\xi_{rad}$\\
$(0.87~R_\odot)$&(g cm$^{-3}$)&($10^6$ K)&(cm$^2$ g$^{-1}$)&(cm$^2$ s$^{-1}$)&(cm$^2$ s$^{-1}$)&(cm$^2$
s$^{-1}$)&(cm$^2$ s$^{-1}$)\\
\\
\hline
\\
$r \simeq 0.7$& $0.4$& $2.6$ & $20$&$15$&$1$&$496$&$4 \times 10^6$\\
$r \simeq 0.3$& $20$ & $7$ & $3.5$&$3.5$&$0.1$&$112$&$1.8 \times 10^5$ \\
\\
\hline
\end{tabular}
\end{center}}
\label{tab:one}
\end{table}%

\begin{table}[htdp]
\caption{Conditions Relevant to Diffusive Stability in the Early Sun's
Radiative Zone} {\footnotesize
\begin{center}
\begin{tabular}{ccccc}
\hline
\hline
\\
Radius & $N^2$&$\Omega_\odot^2 / N^2$&$\epsilon_{\nu}$&$\epsilon_{\eta}$\\
$(0.87~R_\odot)$&(Hz$^2$)& &&\\
\\
\hline
\\
$r \simeq 0.7$& $6 \times 10^{-6}$& $1.2 \times 10^{-6}$&$4 \times 10^{-6}$&$1.2 \times
10^{-4}$\\
$r \simeq 0.3$&$6 \times 10^{-6}$& $1.2 \times 10^{-6}$ &$2 \times 10^{-5}$ & $6.2 \times
10^{-4}$\\
\\
\hline
\end{tabular}
\end{center}}
\label{tab:two}
\end{table}%

\clearpage

\clearpage

\clearpage

\clearpage

\begin{figure}
\plottwo{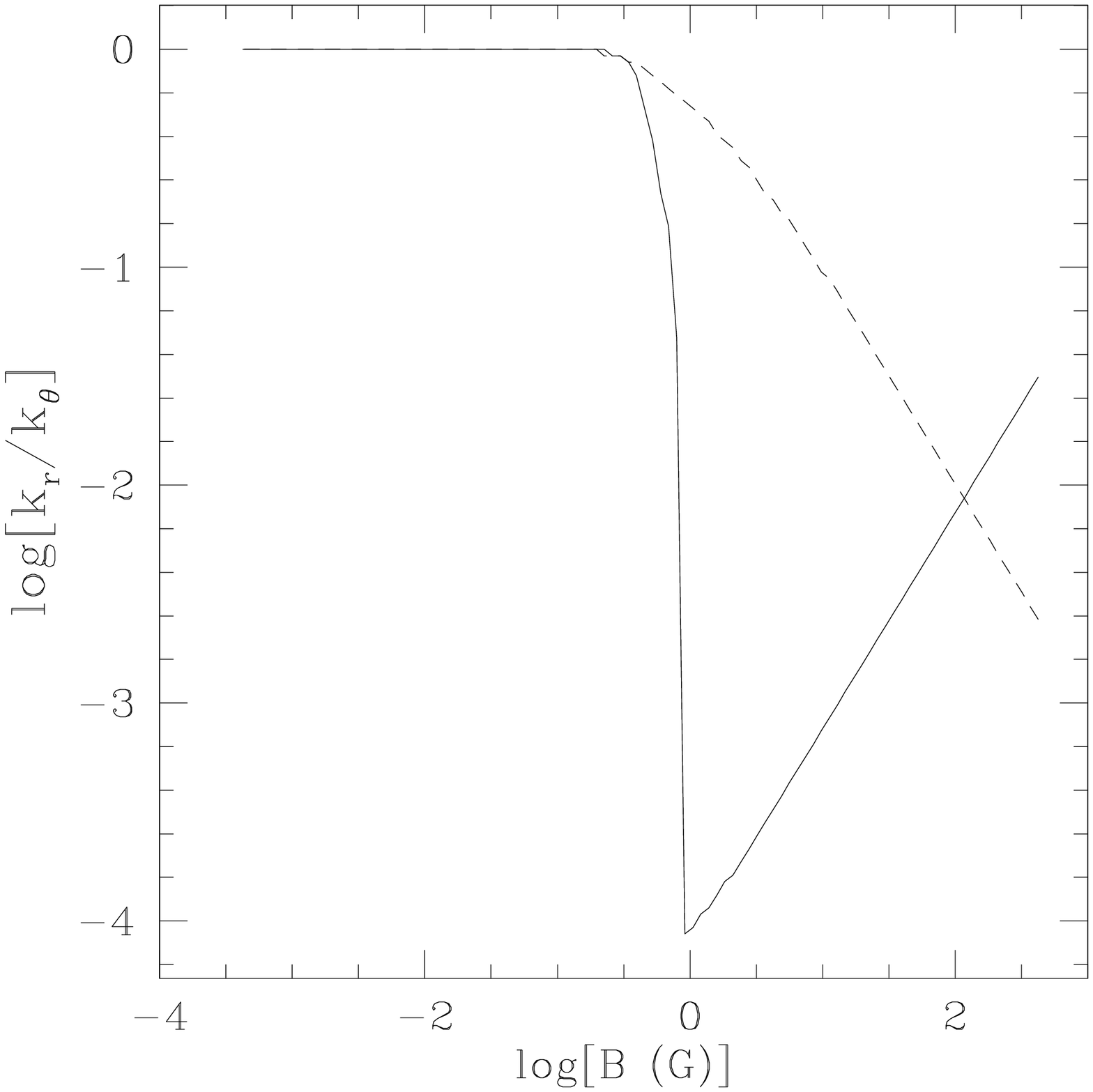}{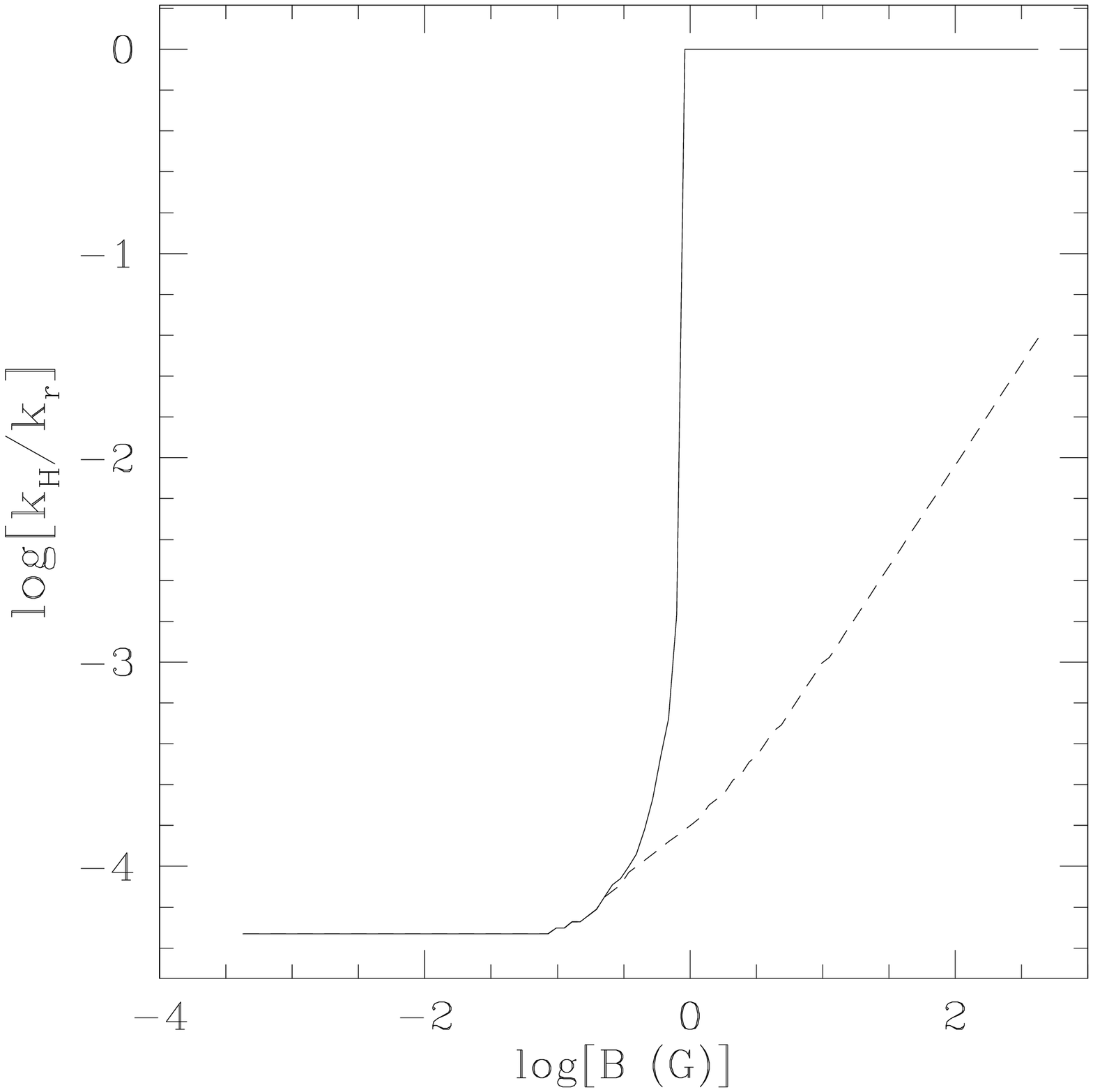}
\caption[]{\label{fig:four} Ratio of radial to polar wavenumbers
(left) and radial wavelength to local scale height (right) for the
same unstable modes as shown in Fig.~\ref{fig:three}. MHD modes (at
large $B$ values) have preferentially radial displacements, with
relatively large radial wavelengths (reaching the local scale height
limit in the polar field case -- solid line).  Hydrodynamical modes
(at low $B$ values) have much smaller wavelengths.}
\end{figure}
\end{document}